\documentclass[trackchanges,twocolumn,twocolappendix]{aastex701}

\begin{document}

\title{The 6 year radio lightcurve of the tidal disruption event AT2019azh}

\author[sname='Burn']{M. Burn}

\affiliation{International Centre for Radio Astronomy Research - Curtin University, GPO Box U1987, Perth WA6845, Australia}
\email{20566566@student.curtin.edu.au}  

\author[sname='Goodwin']{A. J. Goodwin} 

\affiliation{International Centre for Radio Astronomy Research - Curtin University, GPO Box U1987, Perth WA6845, Australia}
\email{}

\author[sname=Anderson]{G. E. Anderson}
\affiliation{International Centre for Radio Astronomy Research - Curtin University, GPO Box U1987, Perth WA6845, Australia}
\email{}

\author[sname=Miller-Jones]{J. C. A. Miller-Jones}
\affiliation{International Centre for Radio Astronomy Research - Curtin University, GPO Box U1987, Perth WA6845, Australia}
\email{}

\author[sname=Cendes]{Y. Cendes}
\affiliation{Department of Physics, Oregon University, 1371 E 13th Ave, Eugene OR 97403, USA}
\email{}

\author[sname=Christy]{C. Christy}
\affiliation{Department of Astronomy and Steward Observatory, University of Arizona, 933 North Cherry Avenue, Tucson, AZ 85721-0065, USA}
\email{}

\author[sname=Lu]{W. Lu}
\affiliation{Department of Astronomy and Theoretical Astrophysics Center, University of California at Berkeley, Berkeley, CA 94720, USA}
\email{}

\author[sname=van Velzen]{S. van Velzen}
\affiliation{Leiden Observatory, Leiden University, Postbus 9513, 2300 RA Leiden, The Netherlands}
\email{}

\begin{abstract}
Analysis of radio emission from tidal disruption events allows for detailed constraints on the properties of ejected outflows and the host environment surrounding the black hole. However, the late-time radio behaviour of tidal disruption events is not well-studied due to a lack of observations. In this work we present long-term radio monitoring observations of the tidal disruption event AT2019azh spanning 1167-2159\,days post disruption. We fit a physically motivated synchrotron model to the radio spectra at each epoch, and model the decay of the light curve under the assumption that the outflow transitions into the non-relativistic Sedov-Taylor regime at late times. Through this modelling we obtain strong constraints on the density profile of the circumnuclear medium, finding an unusually flat density profile proportional to $r^{0.24^{+0.11}_{-0.15}}$. Overall we find that unlike some tidal disruption events, AT2019azh does not show late time re-brightening up to 6\,yr post-disruption. The Sedov-Taylor light curve decay model provides a good fit to the data, motivating the assumption that the outflow was produced by a single ejection of material close to the time of stellar disruption. From forward modelling the evolution of the radio light curve decay we predict that the radio afterglow from AT2019azh should be detectable for decades at its current rate of evolution.

\end{abstract}


\section{Introduction}
\label{sec:intro}
A tidal disruption event (TDE) occurs when a star is destroyed by the tidal forces of a supermassive black hole \citep[SMBH;][]{Rees1988}. Around half of the stellar debris is unbound from the system, while the rest of the material remains bound, producing a luminous flare as it accretes onto the SMBH \citep{Evans1989,Lacy1982}. TDEs provide the opportunity to observe the process of accretion onto a SMBH over a timescale of years, acting as an avenue to watch events such as the evolution of an accretion disk and the launching of outflows over human timescales \citep[for a review, see][]{Gezari2021}.

Optical emission from TDEs is often brighter and more prevalent than theory would suggest, with two main explanations. First is the re-processing envelope, in which an optically thick envelope of debris forms around the black hole. This envelope absorbs the X-rays emitted by the accretion disc and reprocesses them into optical photons \citep{Loeb_1997,Strubbe2009,Metzger2016}. The second mechanism is stream collisions, where the optical emission is produced by the collisions of the orbiting stellar debris stream with itself while it circularises into a nascent accretion disc \citep{Piran_2015,Guo_2025,Lu2020}. Optical/UV observations of TDEs years post-disruption have revealed plateaus in optical emission that can be modelled as the accretion flow of a long-lived accretion disk \citep{Mummery2024}, showing that indicators of a TDE can remain even after the initial flare has faded.   

The X-ray properties of TDEs are varied, as only some TDEs produce visible X-rays and those that do have been seen to produce bright X-ray emission both promptly \citep{Malyali2024} and after a delay \citep{guolo2024}. The presence of delayed and X-ray faint events is thought to be tied to the mechanism of the optical emission, in that a reprocessing envelope obscures the majority of the X-ray emission \citep{Dai2018} at early times, or that production of X-rays is delayed during the stream collision scenario due to the time it takes for the debris to circularise into an accretion disc. 

Radio emission from TDEs probes their outflows, although not all TDEs produce detectable radio emission. TDE radio emission is characterised by synchrotron radiation produced by free electrons in the circumnuclear medium (CNM) being accelerated by collisions with outflows launched by the TDE \citep{Alexander_2020}. The exact mechanisms that launch the outflow are unknown, but in non-relativistic TDEs are generally proposed to be accretion-induced disc winds \citep[e.g.][]{Alexander2016}, collisions of the stellar debris stream \citep{Lu2020}, the unbound debris stream from the initial disruption of the star \citep[e.g][]{Krolik2016}, or jets. Relativistic jets in TDEs are extremely rare ($\sim$1\% of events) \citep[e.g.][]{Andreoni_2022,Zauderer2011}, and more commonly seen are sub-relativistic outflows consistent with lower levels of collimation \citep[e.g.][]{vanVelzen2016}. Radio emission from TDEs is used to examine the properties of both the TDE outflows, and the environment surrounding the SMBH.

Long-term radio monitoring revealed a population of TDEs that produce radio flares at late times \citep[rising radio emission at $\gtrsim$500 days post disruption;][]{Cendes2022,Horesh_2021a,Goodwin_2025}. This includes both re-flaring events with a second radio peak \citep{Goodwin_2025} and flares from TDEs with no previous radio detection \citep{Cendes2022}. The origin of these flares is currently under debate, with explanations such as the delayed launch of outflows from the accretion disk or energy injection into the existing outflow \citep{Cendes_2024}, an off-axis jet becoming visible \citep{Matsumoto2023}, or the outflow encountering a change in the density of the CNM, including the outflow passing the Bondi radius into a flat density profile \citep{Matsumoto_2024} or gaseous clouds in the surrounding medium \citep{Zhuang_2025}. Comparison between TDEs with and without late-time flares could provide insight into the mechanism behind the properties of late time TDE radio emission. 

The radio emission from non-relativistic TDE outflows is modelled as a self-absorbed synchrotron spectrum where the self-absorption frequency is greater than the minimum energy frequency \citep{GranotSari}. If the peak of the spectrum is detected, it is possible to use equipartition analysis \citep{BarniolDuran_2013} to model the outflow \citep[e.g.][]{Goodwin_2022,Cendes_2021,vanVelzen2016}. This provides constraints on properties such as the radius, velocity, and energy of the outflow, as well as the ambient electron density in the surrounding CNM. Because the equipartition model requires a measurement of the synchrotron self-absorption peak, using this analysis at times long after the initial outflow launch when the self-absorption frequency has moved below the observing band is not possible. At late times the outflow will decelerate and become non-relativistic as it sweeps up mass from the CNM. In events with similar outflow physics such as gamma ray bursts \citep{Berger_2004,BarniolDuran2015} and supernova remnants \citep{Stafford_2019}, the decelerating outflow will approach spherical symmetry and the Sedov-Taylor solution can be applied to the non-relativistic outflow \citep{Sironi_2013}. In these events the Sedov-Taylor solution can be used to model the light curve of the radio emission. This model is dependent on the spectral index of the synchrotron spectrum and the host density profile, so it can be used to constrain one of these properties if the other is known.

The CNM density profiles of galaxies are useful in probing the accretion history of their central SMBHs, as different models of accretion predict different density profiles. For example, spherically symmetric Bondi accretion predicts a CNM density profile proportional to $r^{-3/2}$ \citep{Bondi1952}. The density profiles of M87 and NGC 3115 have been constrained using X-ray observations \citep{Russell2015,Wong2011,Wong2014}, but this method requires the central regions of the galaxies to be resolved. TDEs provide an alternative method of probing the CNM density of galaxies. Observations of a TDE can constrain the host CNM density profile through measurements of the CNM electron density using equipartition measurements \citep[e.g.][]{Goodwin2023a,Cendes_2021,Hajela_2025}. This allows radio TDEs to probe the accretion history of their hosts.

In this paper, we present the 1167--2159 day radio light curve of AT2019azh, extending the observed radio light curve of AT2019azh to 6 years. AT2019azh is a TDE that was discovered by the Zwicky Transient Facility (ZTF) \citep{vanVelzen2019} and the All-Sky Automated Survey for Supernovae (ASAS-SN) \citep{Brimacombe2019} on 12 February 2019 and 22 February 2019 respectively. Located at a position (RA, Dec) 08:13:16.95, +22:38:54.02 and with a redshift $z=0.022$, AT2019azh was detected at X-ray, UV, optical \citep{vanVelzen2019}, and radio \citep{PerezTorres2019} wavelengths. \citet{Hinkle2021} found that the optical emission rose consistent with flux $\propto t^2$, and that the X-ray emission was detected around 30 days after the estimated time of first light and brightened at later times.

AT2019azh is a well observed source across the electromagnetic spectrum. \citet{Liu_2022} found that the X-ray and UV/optical emission were not temporally correlated, and suggested that the UV/optical peak was produced by stream-stream collisions during circularisation rather than reprocessing of the X-ray emission. \citet{Faris_2024} presented observations showing a change in the slope and possible bump in the optical light curve of AT2019azh. \citet{Hinkle2021,vanVelzen2021,Liu_2022} all note a late brightening of the X-ray emission around 200 days after discovery. While \citet{Liu_2022} attributed this to delayed accretion, \citet{Hinkle2021} instead interpreted it as an increase in the area of the X-ray emitting region. \citet{Reynolds2025} showed that the IR emission from AT2019azh requires a dense reprocessing envelope, and found that this could be explained by stream-stream collisions or by a viewing angle dependent model \citep{Dai2018}.

At radio wavelengths, \citet{Sfaradi_2022} found that over 400 days the 15.5\,GHz emission plateaued at 100 days, then rose again and peaked around 360 days (see Figure \ref{fig:lightcurve}). They modelled this as a late-time flare from a second ejection of material produced by a transition in the accretion state. We note that this would have occurred at an early time in comparison to other late-time flares \citep[e.g.][]{Cendes_2024}. However, \citet{Goodwin_2022} had previously found that the radio emission from 1.5 - 11\,GHz rose and peaked almost 2 years after the initial disruption, showing light curves consistent with a single long-term rise and decay that included variability that increased with frequency. They favoured a stream-stream collision interpretation, explaining variations in the electron energy index as the result of an inhomogeneous circumnuclear medium, which would also explain the flux density variations seen at higher frequencies in the \citet{Sfaradi_2022} analysis.

The structure of the paper is as follows. In Section \ref{sec:observations} we present an overview of the observations and data processing. In Section \ref{sec:lightcurve} we present the light curve of AT2019azh. In Section \ref{sec:model} we describe the synchrotron model used and the resulting spectral fit for the observations, and we also fit the light curve of AT2019azh to a Sedov-Taylor blast wave model. We discuss the outflow modelling, the host galaxy, compare AT2019azh with other radio emitting TDEs, and present future prospects for long-term observations of radio emission from TDEs in Section \ref{sec:disc}. Finally, we provide a summary in Section \ref{sec:conclusion}.


\section{Observations}
\label{sec:observations}
\subsection{VLA}
Observations of AT2019azh were acquired with the Karl G. Jansky Very Large Array (VLA) between 9 March 2019 and 5 July 2024. In order to track the late-time radio decay of AT2019azh, we observed the source with the VLA 8 times between 20 April 2022 and 5 July 2024. In addition, 13 epochs of observations of the radio rise and peak taken between 9 March 2019 and 5 June 2021 were presented in \cite{Goodwin_2022} and are included in our analysis. In all observations, 3C 147 was used for flux and bandpass calibration. The secondary phase calibrators used were ICRF J083216.0+183212 for 8-12\,GHz (X-band) and 4-8\,GHz (C-band), and ICRF J084205.0+183540 for 2-4\,GHz (S-band) and 1-2\,GHz (L-band). A summary of the new observations presented in this work is given in the appendix with Table \ref{tab:results}.

All data obtained with the VLA were reduced with the Common Astronomy Software Application package version 6.6.1 (\textsc{CASA}, \citet{TheCasaTeam2022}), using the provided VLA calibration pipeline. Images of the field were created using the \textsc{CASA} task \texttt{tclean}, and flux densities were determined by fitting an elliptical Gaussian fixed to the size of the synthesized beam using the task \texttt{imfit}. The flux density uncertainties include the statistical uncertainty determined by the \texttt{imfit} task, and an estimated 5\% of the flux density to account for uncertainty in the flux density bootstrapping, added in quadrature.

\subsection{uGMRT}
We observed AT2019azh on three occasions with the upgraded Giant Metrewave Telescope (uGMRT). Observations were taken in band 5, with a central frequency of 1.26\,GHz and 400\,MHz total bandwidth, and band 4, with a central frequency of 0.65\,GHz and 300\,MHz total bandwidth. Each of the observing bands were broken into 2048 spectral channels. All data were reduced in CASA using standard procedures, including flux and bandpass calibration with 3C 147 and phase calibration with ICRF J084205.0+183540. Observations of the target field were imaged using the CASA task \texttt{tclean} and the flux density of the target was extracted by fitting a Gaussian to the size of the synthesised beam using the CASA task \texttt{imfit}. 

\section{Results}
\label{sec:lightcurve}
\begin{figure}[h]
    \centering
    \includegraphics[width=1\linewidth]{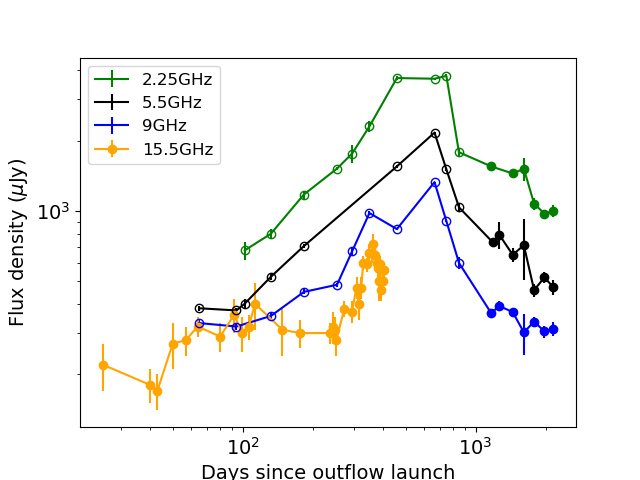}
    \caption{2.25, 5.5, 9, and 15.5\,GHz light curve of AT2019azh. Empty markers are epochs previously presented in \citet{Goodwin_2022}, while filled markers are new observations presented in this work. 15.5\, GHz data taken from \citet{Sfaradi_2022}.}
    \label{fig:lightcurve}
\end{figure}

The light curves of AT2019azh at 2.25, 5.5, 9, and 15.5\,GHz are plotted in Figure \ref{fig:lightcurve}. Over the 1000-2000 days spanned by our new observations, the radio flux density continued to decay, with no current sign of re-flaring that is seen in some TDEs \citep{Cendes_2024,Goodwin_2025}. At all frequencies the radio emission initially decayed steeply post-peak, before flattening at times $>$1000 days post outflow launch, which is most pronounced at 9\,GHz. Both the 5.5\,GHz and 2.25\,GHz light curves appear to be flattening in the last three epochs ($>$1700 days).

\section{Modelling}
\label{sec:model}
\subsection{Spectral Fitting}
\label{sec:spectra}
The radio emission from AT2019azh is consistent with an expanding synchrotron region where the peak of the spectrum is due to synchrotron self-absorption \citep[e.g.][]{Goodwin_2022,Sfaradi_2022}. We model the observed radio spectra and light curves under this assumption to determine properties of the evolving outflow and environment.

We fit the radio spectrum at each epoch using the synchrotron self-absorption model given by \citet{GranotSari} (See appendix \ref{AppendixA} for justification), with the assumption that $\nu_m < \nu_{a} < \nu_c $ (here $\nu_m$ is the minimum synchrotron frequency, $\nu_{a}$ the self-absorption frequency, and $\nu_c$ the synchrotron cooling frequency). The flux density is given by 

\begin{equation}
F_{\nu} = F_{\nu,\text{ext}} \left[ \left(\frac{\nu}{\nu_{a}}\right) ^{-s\beta_1} + \left(\frac{\nu}{\nu_{a}}\right)^{-s\beta_2} 
 \right] ^{-1/s}
\label{eq:specfit}
\end{equation}

\noindent where $F_{\nu,\text{ext}}$ is the normalisation factor, $\nu$ is the frequency, $s = 1.47 - 0.21p$ (from the assumption $k=0$, where the environment density profile is $\propto r^{-k}$ and $r$ is the radius from the black hole), $\beta_1 = 5/2$, $\beta_2 = \frac{1-p}{2}$, and $p$ is the index of the electron energy distribution. 

We fit the spectra using the Python Markov Chain Monte Carlo implementation \texttt{emcee} \citep{MCMC}, using a Gaussian likelihood where the variance is underestimated by a fraction $f$. We assume a uniform prior distribution for our free parameters $F_{\nu,ext}$, $\nu_{a}$, $p$, and $\log(f)$. The prior ranges are given in Table \ref{tab:spectable}. The spectra were fit jointly with a separate $F_{\nu,ext}$ and $\nu_a$ parameter for each epoch, while $p$ and $log_{10}f$ were locked between epochs. This provides a single value of $p$ which fits all epochs. The fit was run using 500 chains for 5000 steps, with the first 2000 steps discarded as burn-in.

\begin{figure*}
    \centering
    \includegraphics[width=0.8\linewidth]{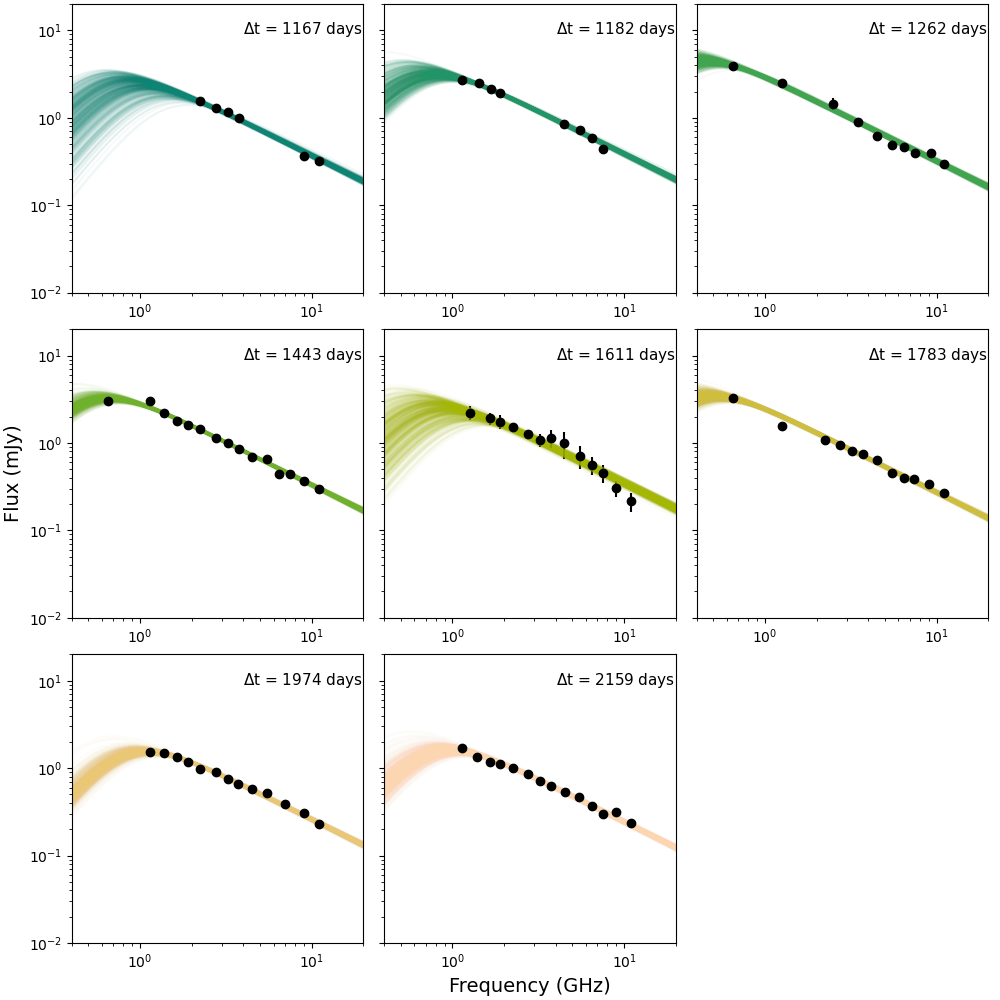}
    \caption{MCMC spectral fits of the radio observations (scatter points) using Equation \ref{eq:specfit} are given by the solid lines. The peak frequency is not well constrained for a majority of the epochs due to a lack of data at low frequencies. $\Delta t$ is time since the launch of the outflow.}
    \label{fig:spectra}
\end{figure*}

Figure \ref{fig:spectra} presents the fitted spectra for each new epoch and the fit parameters are given in Table \ref{tab:spectable}. A lack of coverage at low (sub-GHz) frequencies means that the peak frequencies $\nu_{a}$ and peak flux densities of the spectra are not well constrained by the fit, so we present the peak frequencies as upper limits. The model fits a value of $p=2.95^{+0.03}_{-0.03}$, largely consistent with the values of $p$ at earlier times determined by \citet{Goodwin_2022}, where $p$ was allowed to freely vary between epochs. Given that the synchrotron spectrum used assumes  $\nu_m < \nu_{a} < \nu_c $, it is likely that the synchrotron cooling frequency is $>11$\,GHz, as it is not seen in the frequency range we observed (as tested in appendix \ref{AppendixA2}). 

\begin{table}[h]
    \centering
    \caption{Values of the spectral fit parameters. Parameters free between epochs are $F_{\nu,ext,t}$ and $\nu_{a,t}$ for each epoch, where $t$ is the time since outflow launch. Parameters fixed between epochs are $p$ and $log_{10}f$. Break frequency upper limits are the 99th percentile of the posterior distribution.}
    \begin{tabular}{ccc}
    \hline\hline
        Parameter & Prior & Posterior \\
        \hline
       $F_{\nu,ext,1167}$ (mJy)  & $1$ -- $10^{5}$ & $4.63^{+0.80}_{-0.79}$\\
       $F_{\nu,ext,1182}$ (mJy) & $1$ -- $10^{5}$ & $6.23^{+1.07}_{-0.59}$ \\
       $F_{\nu,ext,1262}$ (mJy) & $1$ -- $10^{5}$ & $9.37^{+1.59}_{-1.09}$\\
       $F_{\nu,ext,1443}$ (mJy) & $1$ -- $10^{5}$ & $6.47^{+0.54}_{-0.38}$\\
        $F_{\nu,ext,1611}$ (mJy)& $1$ -- $10^{5}$ & $4.94^{+1.15}_{-0.89}$ \\
       $F_{\nu,ext,1783}$ (mJy)& $1$ -- $10^{5}$ & $7.04^{+0.94}_{-0.61}$\\
        $F_{\nu,ext,1974}$ (mJy)& $1$ -- $10^{5}$ & $3.12^{+0.24}_{-0.16}$\\
        $F_{\nu,ext,2159}$ (mJy)& $1$ -- $10^{5}$ &  $3.24^{+0.39}_{-0.25}$\\
        $\nu_{a,1167}$ (GHz) & $0.01$ -- $10$ & $  <1.20 $   \\
        $\nu_{a,1182}$ (GHz)& $0.01$ -- $10$ &  $  <0.76 $\\
        $\nu_{a,1262}$ (GHz)& $0.01$ -- $10$ &  $  <0.44 $\\
        $\nu_{a,1443}$ (GHz)& $0.01$ -- $10$ & $  <0.57 $ \\
        $\nu_{a,1611}$ (GHz)& $0.01$ -- $10$ &  $ <1.06 $\\
        $\nu_{a,1783}$ (GHz)& $0.01$ -- $10$ & $ <0.46 $ \\
        $\nu_{a,1974}$ (GHz)& $0.01$ -- $10$ & $ <0.92 $ \\
        $\nu_{a,2159}$ (GHz)& $0.01$ -- $10$ & $ <0.89$ \\
        $p$  & $1$ -- $4$ & $2.95^{+0.03}_{-0.03}$ \\
        $log_{10}f$ & $-10$ -- $10$ & $-3.02^{+0.16}_{-0.15}$\\
        \hline
    \end{tabular}
    \label{tab:spectable}
\end{table}

\subsection{Light Curve Model}
\label{sec:ST}
For a freely expanding synchrotron-emitting source, we expect that the light curve will evolve as an expanding self-absorbed synchrotron emitting region before the self-absorption peak, due to the peak frequency of the radio spectrum decreasing with time and increasing outflow radius \citep[as captured by][]{Goodwin_2022}. After the synchrotron self-absorption peak passes a given frequency, the light curve is expected to decay as the outflow continues to expand and begins to decelerate after picking up sufficient mass from the CNM. In this section, we model the light curve under the assumption that the outflow is an adiabatically expanding blast wave governed by the Sedov-Taylor solution and will decelerate once it has swept up mass similar to its own \citep{Lu2020}. In this model the radio emission is produced from the outflow acting as a spherical shock impacting a stratified medium with a density profile $r^{-k}$, where $r$ is the radius. Using the Sedov-Taylor blast wave to model the light curve of AT2019azh is motivated by \citet{Goodwin_2022}, who favoured a single collision-induced outflow as the origin of the radio emission. The outflow radius and velocity measured by \citet{Goodwin_2022} from equipartition analysis appear to support this interpretation, with a generally constant outflow velocity and linear radius increase until the last epoch, just after the peak of the light curve. 

In a single ejection scenario the Sedov-Taylor model can be used to constrain the value of $k$ after the peak of the outflow. This constraint is an independent method to determine the host environment to that of the ambient electron densities obtained through the equipartition method commonly used in TDE outflow analysis. It can probe the density profile at large radii where the peak frequency of the spectra has fallen below the observed frequencies.

The expected evolution of the flux density in the Sedov-Taylor phase is given by \citet{Sironi_2013}. We adapt the equations from \citet{Goodwin_2025} to develop an equation to describe the evolution of the flux density in the Sedov-Taylor phase, given by

\begin{equation}
    \label{eq:st1}
    F_\nu \propto  \nu^{\frac{1-p}{2}} \left(t\right) ^{-\frac{3(p+1)}{2(5-k)}} \text{ for } 2<p<3,
\end{equation}
\begin{equation}
    \label{eq:st2}
    F_\nu \propto  \nu^{\frac{1-p}{2}} \left(t\right) ^{-\frac{2(3-k)(p-3)+3(p+1)}{2(5-k)}} \text{ for } p>3,
\end{equation}

\noindent where $t$ is the current time since the outflow launch. 

We use Equation \ref{eq:st1} as we found $p < 3$ in Section \ref{sec:spectra}. While we assumed that $k=0$ in Equation \ref{eq:specfit}, only the sharpness of the self-absorption break depends on this assumption. As the self-absorption frequency was unconstrained by our observations, we assume that the self-absorption break has negligible influence on the slope of the spectra at the frequencies covered by our observations.

We combine the Sedov-Taylor model for the light curve decay with a power law $At^{\alpha}$ to capture the rise of the light curve into the form of a smoothly broken power law to create the model formula for the light curve, given by

\begin{equation}
    \label{eq:lcmodel}
    F_\nu = A \nu^{\frac{1-p}{2}}\left[(t/t_b)^{-s\alpha} +(t/t_b)^{-s\frac{-3(p+1)}{2(5-k)}}\right]^{-1/s}
\end{equation}

\noindent where $A$ is a normalisation factor, $s$ is the break sharpness, $t_b$ is the time of the peak flux density, and $\alpha$ is the index of the rise of the light curve.

We fit Equation \ref{eq:lcmodel} to the individual frequency light curves using \texttt{emcee}, using 200 walkers of 5000 steps, discarding the first 2000 as burn-in. We use uniform priors for $A$, $k$, $t_b$, $s$, $\alpha$, and $\text{log}(f)$, presented in Table \ref{table:params}. The prior range chosen for $t_b$ is dependent on the frequency, and is the time between the two epochs directly before and after the epoch with peak flux density (e.g. 459 to 749 days at 9\,GHz). Individual values of $A$ and $t_b$ are fit for each frequency, while the other parameters are fixed between all frequencies as they are frequency independent.

\begin{figure*}
    \centering
    \includegraphics[width=1\linewidth]{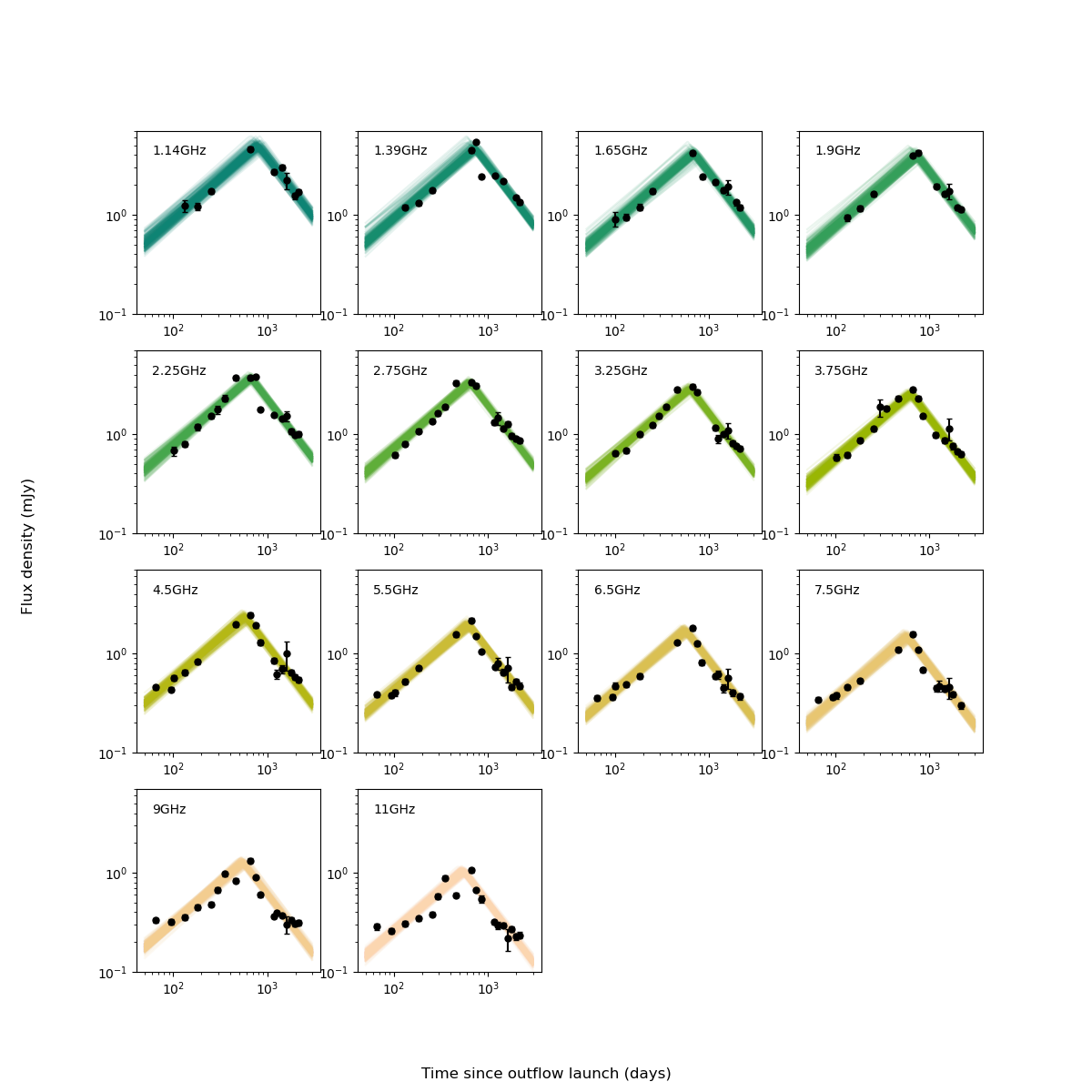}
    \caption{Individual frequency light curve fits for the Sedov-Taylor model generated by MCMC modelling using Equation \ref{eq:lcmodel}. The range of fits are broadly consistent with the observed flux densities, but with some deviation due to more structured light curves at 9 and 11\,GHz.}
    \label{fig:seplight}
\end{figure*}

Figure \ref{fig:seplight} shows all MCMC generated fits for each frequency. Generally, the observed flux densities at lower frequencies are within the range of fits generated by the MCMC model. However, at the highest frequencies there is variability in the rate of the rise and at late times there is a more pronounced flattening of the light curve, resulting in a larger number of data points falling outside of the range of the MCMC fits compared to the lower frequencies. From Table \ref{table:params}, the fit finds a flat density profile with $k=0.24^{+0.11}_{-0.15}$, and the peak times at each frequency, $t_{b,\nu}$, overall follow the trend of decreasing as the frequency increases. This frequency-dependent behaviour of the peak times is expected for an expanding synchrotron self-absorbed region.

\begin{table}[h]
    \centering
    \caption{Values of the model light curve parameters determined by the fit, which are the normalisation factor $A$ and time of peak $t_b$ for each frequency, the density profile index $k$, the break sharpness $s$, temporal index of the rise $\alpha$, and the log of the variance underestimation fraction $f$.}
    \begin{tabular}{ccc}
    \hline\hline
        Parameter & Prior & Posterior \\
        \hline
        $A_{1.13}$ (mJy/GHz)& $1$ -- $10^{4}$ &  $5.65^{+0.41}_{-0.40}$  \\
         $A_{1.39}$ (mJy/GHz)& $1$ -- $10^{4}$& $6.55^{+0.42}_{-0.42}$  \\
         $A_{1.65}$ (mJy/GHz)& $1$ -- $10^{4}$ & $6.60^{+0.43}_{-0.37}$ \\
         $A_{1.9}$ (mJy/GHz)&$1$ -- $10^{4}$& $7.40^{+0.43}_{-0.47}$\\
         $A_{2.25}$ (mJy/GHz)&$1$ -- $10^{4}$&  $8.16^{+0.33}_{-0.34}$  \\
         $A_{2.75}$ (mJy/GHz)&$1$ -- $10^{4}$ & $8.94^{+0.46}_{-0.50}$  \\
         $A_{3.25}$ (mJy/GHz)& $1$ -- $10^{4}$& $9.16^{+0.51}_{-0.42}$  \\
        $A_{3.75}$ (mJy/GHz)&$1$ -- $10^{4}$& $9.67^{+0.62}_{-0.54}$ \\
         $A_{4.5}$ (mJy/GHz)&$1$ -- $10^{4}$& $10.2^{+0.65}_{-0.59}$ \\
         $A_{5.5}$ (mJy/GHz)&$1$ -- $10^{4}$& $10.4^{+0.56}_{-0.56}$\\
        $A_{6.5}$ (mJy/GHz)& $1$ -- $10^{4}$& $10.3^{+0.54}_{-0.55}$\\
        $A_{7.5}$ (mJy/GHz)&$1$ -- $10^{4}$&  $10.8^{+0.57}_{-0.66}$\\
        $A_{9}$ (mJy/GHz)&$1$ -- $10^{4}$&   $11.0^{+0.65}_{-0.48}$ \\
        $A_{11}$ (mJy/GHz)&$1$ -- $10^{4}$&  $10.9^{+0.51}_{-0.58}$ \\
        $t_{b,1.13}$ (days)&$459$ -- $1182$&  $776^{+49}_{-54}$ \\
        $t_{b,1.39}$ (days)&$459$ -- $849$&  $735^{+38}_{-41}$ \\
        $t_{b,1.65}$ (days)& $459$ -- $849$&  $728^{+37}_{-39}$ \\
        $t_{1.9}$ (days)& $459$ -- $1182$& $753^{+49}_{-39}$ \\
        $t_{b,2.25}$ (days)& $459$ -- $749$& $666^{+35}_{-25}$  \\
        $t_{b,2.75}$ (days)&$459$ -- $749$&  $674^{+34}_{-33}$ \\
        $t_{b,3.25}$ (days)&$459$ -- $749$&  $632^{+24}_{-25}$ \\
        $t_{b,3.75}$ (days)&$459$ -- $749$&  $633^{+29}_{-33}$ \\
        $t_{b,4.5}$ (days)&$459$ -- $749$&  $588^{+31}_{-23}$  \\
        $t_{b,5.5}$ (days)&$459$ -- $749$&  $599^{+31}_{-31}$ \\
        $t_{b,6.5}$ (days)&$459$ -- $749$&  $575^{+31}_{-28}$  \\
        $t_{b,7.5}$ (days)&$459$ -- $749$&  $556^{+27}_{-31}$  \\
        $t_{b,9}$ (days)&$459$ -- $749$& $563^{+27}_{-37}$ \\
        $t_{b,11}$ (days)&$459$ -- $749$& $556^{+27}_{-30}$\\
        $k$&$0$ -- $4$&  $0.24^{+0.11}_{-0.15}$  \\
        $s$& $1$ -- $100$& $53.9^{+29.4}_{-26.4}$  \\
        $\alpha$&$0.01$ -- $5$&  $0.82^{+0.03}_{-0.02}$  \\
        $log_{f}$&$-10 $ -- $ 10$& $-1.77^{+0.06}_{-0.06}$ \\
        
       \hline
    \end{tabular}
    \label{table:params}
\end{table}

\section{Discussion}
\label{sec:disc}

The radio observations we present of AT2019azh from 1167 to 2159 days post outflow launch show a decline in the radio flux density at all frequencies. This decline continues the decay that began after the peak of the radio light curve was reached at $\sim666$ days, captured by \citet{Goodwin_2022}. As of our final observation 2159\,days post outflow launch, AT2019azh does not show a second rise in radio emission like those seen in $\sim$40\% of TDEs \citep{Cendes_2024} such as ASASSN-15oi \citep{Horesh_2021a} and AT2020vwl \citep{Goodwin_2025}.

\subsection{The outflow of AT2019azh}
\label{sec:outflow}

The nature of the radio-emitting outflow from AT2019azh has been previously debated. \citet{Goodwin_2022} found the outflow consistent with a single blast wave ejected at a constant velocity likely launched at the time of stellar disruption from stream-stream collisions. \citet{Sfaradi_2022} proposed a second ejection of material, seen in the 15.5\,GHz light curve in Figure \ref{fig:lightcurve}. We found that the Sedov-Taylor model provides a good fit to the radio light curve decay of AT2019azh across the observed frequencies (Figure \ref{fig:seplight}), particularly at the lower frequencies. This suggests that the outflow from AT2019azh is broadly consistent with a single ejection of material moving at initially mildly relativistic speeds, then transitioning to a Newtonian outflow post radio peak.

The joint fitting of the observed radio spectra returns an electron energy distribution index of $p\approx2.95^{+0.03}_{-0.03}$, which is similar to other non-relativistic TDEs \citep{Alexander2016,Cendes_2021,Goodwin2023a}. The Sedov-Taylor model constrains a very flat host galaxy density profile of $r^{-0.24^{+0.11}_{-0.15}}$, which is unusual when compared with the much steeper inferred ambient densities of other TDE environments \citep{Cendes_2021,Cendes2022,Alexander2016}.

The slow evolution of the outflow of AT2019azh supports the flat CNM density profile constrained by the model. The temporal index of the rise of the light curve is dependent on $k$ \citep{GAO2013}, with a smaller value of $k$ resulting in a smaller temporal index. For a relativistic spherical outflow expanding into a medium with $k=0$, \citet{GAO2013} derive the temporal index for the spectral regime $\nu<\nu_m<\nu_{a}$ as $1/2$ and $5/4$ for the regime $\nu_m<\nu<\nu_{a}$. Our fit constrained a value of $0.82^{+0.03}_{-0.02}$, falling between these values, which may suggest that during the rise of the light curve the minimum energy frequency is moving through the observed frequencies. 

\citet{Goodwin_2022} found fluctuations in the outflow energy and the value of $p$ they determined steepened from $p\approx2.6$ to $p=3-3.5$ in two of their spectra, which they suggested was due to changes in the shock acceleration efficiency due to an inhomogeneous CNM. This could explain the varying rate of the rise in the 9 and 11\,GHz light curve shown in Figure \ref{fig:seplight}, with the variability at higher frequencies due to interactions between the outflow and regions of higher CNM density producing changes in the synchrotron emission of smaller groups of electrons.

\citet{Sfaradi_2022} proposed that the radio emission from AT2019azh is produced by two outflows, with the second outflow only ejected after $\sim270$ days. This would explain the changes in the rate of the rise at 9 and 11\,GHz, but the same behaviour is not seen at the lower frequencies. The continued rise of the radio emission past $400$ days indicates that a third ejection is required to explain the light curve in this scenario. This interpretation of the radio emission is incompatible with our Sedov-Taylor model, which has assumed only a single outflow was ejected. We suggest that the consistency of the Sedov-Taylor fit to the long term light curve indicates that AT2019azh has only ejected a single outflow, and the variability in the 15.5\,GHz light curve could be attributed to the clumpy CNM interpretation from \citet{Goodwin_2022}.

\begin{figure}[h]
    \centering
    \includegraphics[width=1\linewidth]{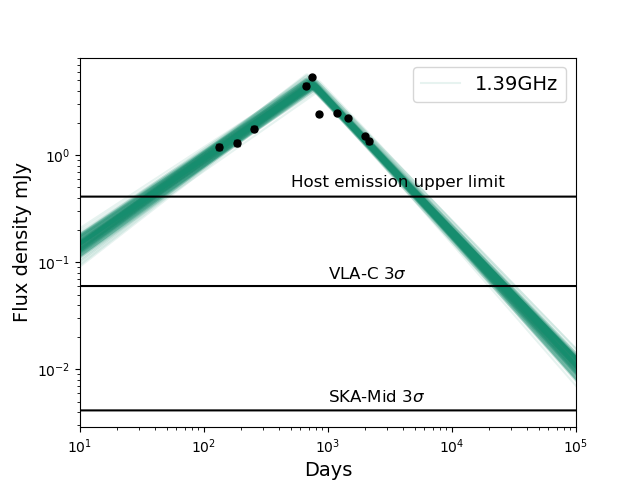}
    \caption{Expected evolution of the 1.39\,GHz radio emission of AT2019azh assuming the Sedov-Taylor decay model (blue line) and the observed 1.39\,GHz flux densities of AT2019azh to date (black points). Included are the 3$\sigma$ sensitivities of the VLA-C configuration and SKA-Mid band2 (0.95-1.76\,GHz) for a 1 hour observation, as well a 1.4\,GHz upper limit of the host galaxy emission from archival Faint Images of the Radio Sky at Twenty centimeters \citep[FIRST;][]{Becker1995}.}
    \label{fig:futureexpect}
\end{figure}

The 6 year radio light curve presented in this work demonstrates that radio afterglow from TDEs is long-lived. Figure \ref{fig:futureexpect} shows the predicted evolution by our Sedov-Taylor fit of the 1.4\,GHz flux density with upper limits for the VLA-C configuration, the upcoming SKA-Mid at 0.95-1.76\,GHz, and an upper limit on AT2019azh's host galaxy emission at 1.4\,GHz from archival Faint Image of the Radio Sky at Twenty centimetres \citep[FIRST;][]{Becker1995}. Following the model the emission should remain above the host galaxy upper limits up to $\sim$5000 days post disruption, and assuming it doesn't fall below the host emission, the TDE should be detectable by the VLA for decades.

\subsection{Environment of the host galaxy}
\label{sec:host}

\begin{figure*}[ht]
    \centering
    \includegraphics[width=1\linewidth]{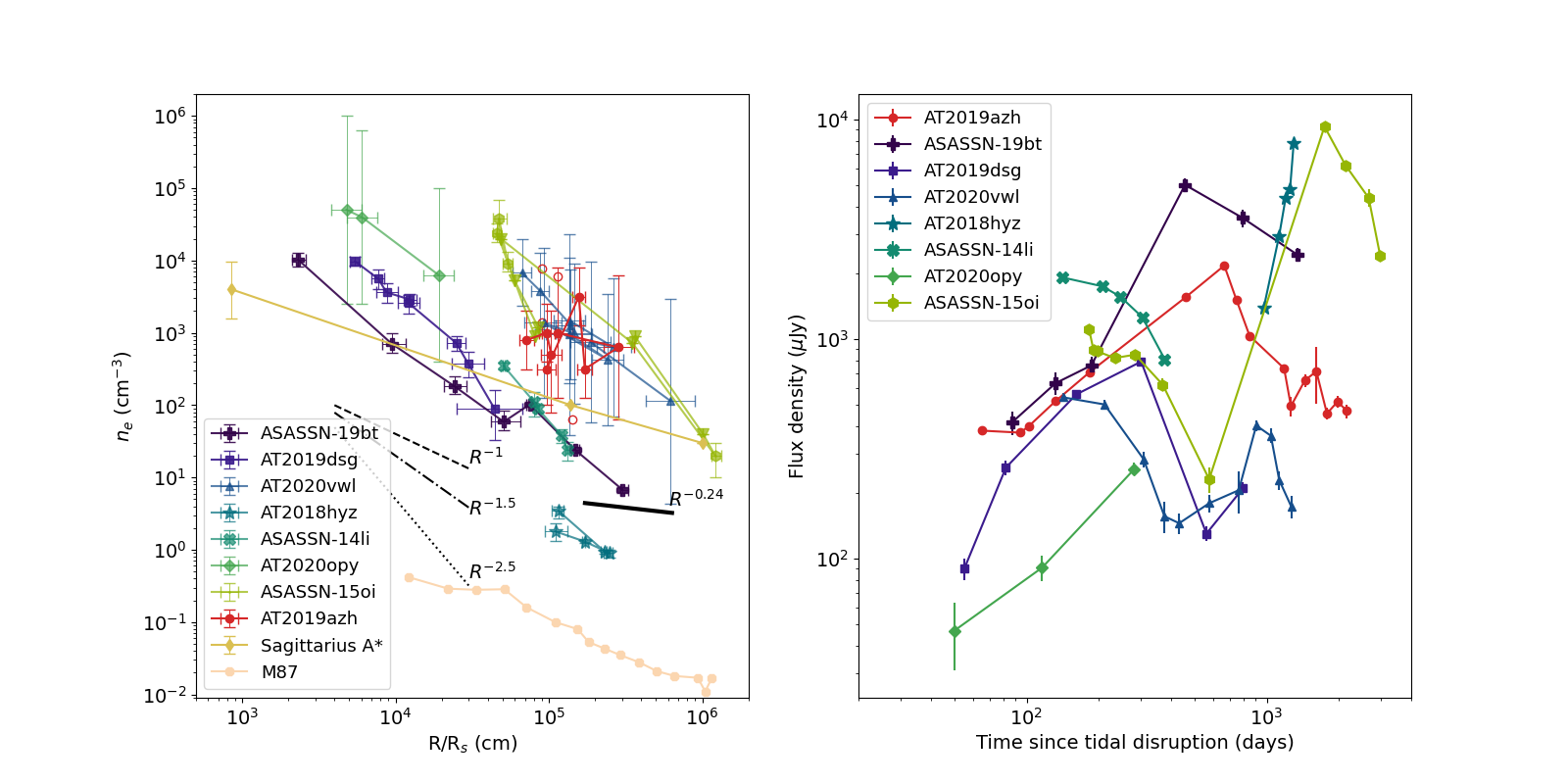}
    \caption{\textit{Left:} The ambient CNM density and the scaled outflow radius inferred from equipartition analysis for multiple radio TDEs, taken from the literature. Also included are M87 and Sagittarius A*. Included as dotted and dashed lines are some example scaling relations. The thick black line is the density profile of AT2019azh constrained in this work. Empty points indicate observations with radio spectra with peaks that are not well-constrained. Downwards triangles for ASASSN-15oi indicate observations with only upper limits on the ambient density and lower limits on the radius.  \textit{Right:} The 5.5\,GHz flux density and time since stellar disruption for a number of radio TDEs published in the literature. Data taken from \citet{Christy_2024} (ASASSN-19bt), \citet{Goodwin_2022,Goodwin2023a,Goodwin_2025,Goodwin2023b} (AT2019azh,AT2020vwl,AT2020opy), \citet{Cendes_2021,Cendes_2024,Cendes2022} (AT2019dsg, AT2018hyz), \citet{Alexander2016} (ASASSN-14li), \citet{Hajela_2025} (ASASSN-15oi),
     \citet{Russell2015} (M87), \citet{Baganoff_2003,Gillessen_2019} (Sagittarius A*).}
    \label{fig:compare}
\end{figure*}

The shallow circumnuclear medium density profile for the host galaxy of AT2019azh we constrained using the Sedov-Taylor fit is unexpected, being flat compared to other constrained SMBH host environments. The left panel of Figure \ref{fig:compare} shows radii for the emitting regions inferred from equipartition analysis and ambient CNM densities published in the literature for TDEs. The thick black line is the density profile we have constrained in this work. The range it is plotted over is an estimated radius of the outflow from 666 -- 2159 days. The lower limit is from the equipartition analysis of \citet{Goodwin_2022} and the upper limit assumes a constant outflow velocity of 0.1$c$, similar to earlier outflow velocities obtained by \citet{Goodwin_2022}. This upper limit is likely an overestimation, since the Sedov-Taylor model assumes the outflow decelerates. The ambient electron densities of some of the TDEs, including AT2019azh, have very large errors that encompass a large range of densities due to the peak of the radio spectrum not being well-constrained by the observations. The size of these errors prevent the density profile from being constrained. In this case the value of $k$ we obtained via modelling the light curve with our Sedov-Taylor model provides a more precise constraint on the CNM density profile than the inferred ambient densities.

For the TDEs that have well constrained ambient electron densities, the density profiles fall in the range $r^{-1}$ - $r^{-2.5}$. Additionally, the SMBH environments of M87, Sagittarius A*, and NGC 3115 are all consistent with a profile of $r^{-1}$ \citep{Russell2015,Baganoff_2003,Gillessen_2019,Wong2011,Wong2014}, meaning that the density profile of AT2019azh seems to be unusual even compared to non-TDE host SMBHs. The flat profile of AT2019azh's host may suggest that the host SMBH has been very inactive in its past and has not accreted much material from the surrounding medium. Comparatively, accreting at the Bondi rate would suggest a density profile of $r^{-3/2}$, and simulations of hot accretion flows find a density profile range of $r^{-0.5}$ -- $r^{-1}$ \citep{Yuan2012}.

\citet{French2023} found that the host galaxy of AT2019azh hosts an extended emission-line region (EELR), suggesting the presence of an ionising source in the galaxy's recent past such as AGN activity or a previous TDE. While AGN activity is the most likely and preferred explanation of \citet{French2023}, this conflicts with the shallow density profile we measure for AT2019azh. This could point towards a previous TDE as the source of the EELR, as the accreted material would then not have to come from the CNM. However the probability of this is quite low, as measurements of TDE rates estimate rates on the order of $10^{-4}$\,galaxy$^{-1}$\,yr$^{-1}$ \citep{Velzen_2018}, while \citet{Wevers_2024} find that the source of the EELR faded within the last $\sim1000$ years.

The host galaxies of the radio-emitting TDEs ASASSN-14li and iPTF-16fnl both also contain an EELR \citep{Wevers_2024}. Archival data of ASASSN-14li suggests weak AGN activity \citep{Alexander2016,vanVelzen2016} in the host galaxy, suggesting that  AGN activity is the cause of the EELR. \citet{Alexander2016} constrain ASASSN-14li's host CNM density profile as $r^{-2.5}$, much steeper than that of the host of AT2019azh. For iPTF-16fnl, \citet{Horesh_2021b} model a CNM density of $r^{-1.3}$ for the likely case of a delayed outflow launched at $\geq$ 56 days. This difference in CNM density profiles may suggest that the source of the EELR is not a defining factor in the formation of the CNM density profile. However, this does not account for the sample size of only three galaxies being too small to make a statistically significant conclusion. It also doesn't factor in the possibility that the EELR in AT2019azh could be the result of a different ionising source such as a previous TDE. 

\subsection{Comparison of TDE light curves}
\label{sec:compare}

The right side of Figure \ref{fig:compare} presents the 5.5\,GHz flux density of AT2019azh compared to other radio-emitting TDEs from the literature. AT2019azh takes noticeably longer to rise than ASASSN-14li and AT2019dsg, both of which began to decay hundreds of days before AT2019azh reached its peak. AT2020vwl's first flare peaks at similarly early times, and while its second flare peaks later it also rises and decays much faster than AT2019azh. Of the two remaining sources, AT2018hyz was not detected in the radio until around 970 days after discovery \citep{Cendes2022}, after which it has very quickly risen. Like AT2019azh, AT2020opy was detected in the radio early, but a lack of observations at later times makes it difficult to tell if it had a similar long rise to peak. \citet{Hajela_2025} captured the $182-2970$\,day radio light curve of ASASSN-15oi, presenting a longer-sampled light curve than AT2019azh. The light curve of ASASSN-15oi captures two flares, with the first appearing to peak and decay on a similar time scale to AT2020vwl's first flare and ASASSN-14li. The second flare was only observed after it began decaying. Even after decaying for 1229 days between its first and last observation, the second flare is still brighter than the first. The light curve of ASASSN-19bt up to around 200 days is similar to AT2019azh, but afterwards its flux density rises faster. ASASSN-19bt also peaks around 200 days before AT2019azh with a higher flux density.

The slower evolution of AT2019azh compared to these other TDEs may be another indicator of it having a comparatively flatter density profile due to the relation between $k$ and the rise time discussed in Section \ref{sec:outflow}. This would suggest that despite the large error in the inferred ambient densities, AT2020vwl may be estimated to have a similar density profile to ASASSN-14li considering their light curves appear to peak around similar times. While the peak of AT2020opy was not captured, the value $\alpha$ for a rise $\propto t^{\alpha}$ between the second and third observations is $\sim1.17$, indicating a steeper CNM density profile than AT2019azh's $\alpha=0.82^{+0.03}_{-0.02}$ despite the non-constraining inferred ambient densities. While ASASSN-15oi's earlier peak in its first flare indicates a steeper density profile, its second flare has been decaying for over 1229 days and is still brighter than the first. This slow evolution could be attributed to the flare being much brighter, but it could also indicate that assuming a single ejection of material, the outflow has encountered a flatter ambient density profile. While the similarity between the early-time light curves of ASASSN-19bt and AT2019azh could suggest a similar density profile, the rate of the rise in ASASSN-19bt increases. \citet{Christy_2024} suggest that the CNM of ASASSN-19bt has a shallower density profile close to the black hole and a steeper profile further out, consistent with the increase in the rate of the light curve rise. 

\subsection{Estimation of the total outflow energy}
Another advantage of the Sedov-Taylor model is that the energy of the blast wave can be estimated. This is because in the Sedov-Taylor solution the evolution of the blast wave is governed by the energy of the initial blast. Using equation 12 of \citet{Sironi_2013}, we estimated the energy of the outflow using our model fit flux densities, taking the value averaged across all frequency fits. We found an outflow energy of $E \approx10^{50} (\frac{n_e}{10})^{\frac{3+3p}{20}}$\,erg, where $n_e$ is the number density of the CNM in units of cm$^{3}$. For a CNM with density $10^4$\,cm$^{-3}$, which is the upper end of the densities for AT2019azh in Figure \ref{fig:compare}, this would give a total outflow energy of $\sim6 \times10^{51}$\,erg. While not an unreasonable value, this is more than the highest value of $10^{50.6\pm0.1}$\,erg inferred through equipartition analysis \citep{Goodwin_2022}. This discrepancy may be the result of the assumed values of uncertain parameters in the equipartition analysis, such as $\epsilon_e$ and $\epsilon_B$.

\section{Conclusions}
\label{sec:conclusion}
We observed the radio evolution of AT2019azh from $\sim$1000-2000 days post disruption. The radio emission has continued to decay at all frequencies after its peak at $\sim650$ days, reported by \citet{Goodwin_2022}, showing no sign of the re-flaring at late times that has been seen in a number of TDEs. Assuming a self-absorbed synchrotron model we jointly fit the spectra of each epoch, finding an overall electron energy index of $p=2.95^{+0.03}_{-0.03}$, consistent with the range of $p$ values \citet{Goodwin_2022} determined during the rise of AT2019azh.

We modelled the light curve evolution of AT2019azh under the assumption of a Sedov-Taylor blast wave, in which a spherical shockwave expands at constant velocity until it sweeps up material from the CNM equal to its mass, after which it begins decelerating. From this model we constrained a very flat CNM density profile of $r^{-0.24^{+0.11}_{-0.15}}$, a result that appears consistent with inferred ambient CNM densities inferred via equipartition analysis of the rise \citep{Goodwin_2022}. This is an unexpected result when compared to other CNM densities constrained by TDEs, which are generally found to have CNM density profiles in the range of $r^{-1}$ to $r^{-2.5}$. Additionally, other active and quiescent galactic nuclei such as M87, Sagittarius A*, and NGC 3115 have steeper CNM density profiles constrained to $r^{-1}$. The temporal evolution of AT2019azh is slower than other TDEs with detailed radio observations in the literature, which indicates a lower value of $k$ in an expanding outflow model. We suggest these factors support the inhomogeneous CNM interpretation for AT2019azh presented by \cite{Goodwin_2022}, with the rather flat CNM density profile of the host galaxy resulting from a lack of significant accretion from the CNM onto the SMBH in the past. The presence of an extended emission line region within the galaxy \citep{French2023,Wevers_2024} could challenge this interpretation if it was created by AGN activity due to the material from the SMBH environment required to power the AGN. A previous TDE could also explain the presence of the EELR and does not require the accreted material to come from the CNM.

In the future more sensitive telescopes such as the SKA will be viable options to observe the long-term radio emission of even faint TDE outflows. Observations of a greater number of TDEs similar to AT2019azh with no late time flares or re-flaring events will help to determine if the Sedov-Taylor solution is an accurate model of TDE outflow evolution. If so, the Sedov-Taylor model provides a method independent and alternative to equipartition analysis to constrain the density profile of the CNM, potentially allowing for greater insight into the population of TDE host galaxies and their interactions with their SMBH.

\begin{acknowledgments}
AJG is grateful for support from the Forrest Research Foundation. This work is supported by an Australian Government Research Training Program (RTP) Scholarship.

\end{acknowledgments}

\facilities{VLA, uGMRT}
\software{CASA \citep{TheCasaTeam2022}, emcee \citep{MCMC}, astropy \citep{Astropy}}

\appendix

\section{Choice of spectral fit model}

\subsection{Physically motivated model}
\label{AppendixA}

\begin{figure}[h]
    \centering
    \includegraphics[width=1\linewidth]{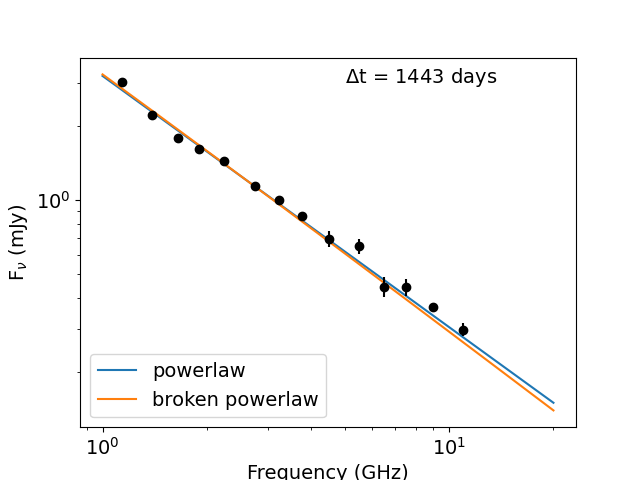}
    \caption{Simple powerlaw and \texttt{astropy} \textsc{SmoothlyBrokenPowerLaw1D} function fit to 1443 day spectra.}
    \label{fig:powerlaw}
\end{figure}

We motivate fitting with the broken power law model of \citet{GranotSari} based on the curvature seen in some epochs not being well-fit by a single power law (e.g. 1611 days). To test the suitability of a broken power law, we fit both a single power law and \texttt{astropy}'s \citep{Astropy} \textsc{SmoothlyBrokenPowerLaw1D} function to the spectra. We find that for spectra well-described by the single power law, the broken power law will give an equally good fit to the spectra, shown in Figure \ref{fig:powerlaw}. The value of $p$ determined from the single power law fit is within error of the value $p$ determined from the broken power law fit. Therefore for consistency we choose to fit all epochs with the physically motivated broken power law model despite the peak of the spectrum being outside of the observed frequency bands post the radio peak.

\subsection{Testing for cooling break}
\label{AppendixA2}
As we are observing the late-time evolution of the outflow, we determine whether the cooling break of the spectra has moved through the observed frequencies. To this end, we compare the fits of the synchrotron model with a single break at the self-absorption frequency $\nu_a$ with a two break model, in which the cooling break frequency $\nu_c$ is the second break. The double-break model is taken from \citep{GranotSari} equation 4 and is given by

\begin{equation}
\label{eq:doublebreak}
    F_{v,2}=F_v \left[1 + \left(\frac{\nu}{\nu_c}\right)^{s_2(\beta_2-\beta_3)} \right]^{-1/s_2}  
\end{equation}

Here $F_v$ is Equation \ref{eq:specfit}, $\beta_2=\frac{1-p}{2}$, $\beta_3=\frac{-p}{2}$, $s_2=1.15-0.06p$ (assuming $k=0$), and $\nu_c$ is the cooling break frequency.

To determine which is the better fit for the data, we compare the Akaike information criterion (AIC) and Bayesian information criterion (BIC) of the two models. The AIC and BIC are

\begin{equation}
    \label{eq:AIC}
    AIC=-2L + 2q
\end{equation}

\begin{equation}
    \label{eq:BIC}
    BIC=-2L + qln(N)
\end{equation}

\noindent here $L$ is the log-likelihood of the model, $q$ is the number of free parameters, and $N$ is the number of data points. 
A model is a statistically better fit to the data if the AIC or BIC are 2 or more points lower than the alternate model. Table \ref{tab:Breakfits} shows that the single-break model is a better fit to the data, with both the AIC and BIC significantly lower than their counterparts determined for the double-break model.

\begin{table}[h]
    \centering
    \caption{Calculated AIC and BIC values for the single break self-absorption model and the double break self-absorption and cooling break model.}
    \begin{tabular}{ccc}
         & Single-break & Double-break \\
         \hline \hline
      AIC  & -371.61 & -318.16  \\
      BIC  & -326.62 & -253.17 \\
         \hline
    \end{tabular}
    \label{tab:Breakfits}
\end{table}

\section{Flux density corrections}

\begin{figure}[h]
    \centering
    \includegraphics[width=1\linewidth]{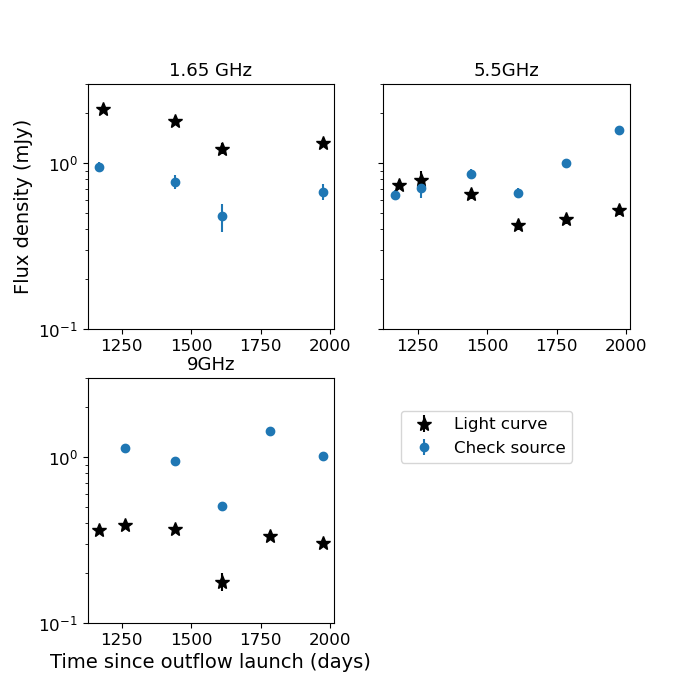}
    \caption{Comparison of measured AT2019azh flux density with measured flux density of a background source in the image field at 1.65, 5.5, and 9\,GHz. Black stars are AT2019azh data and blue dots are the background source.}
    \label{fig:checksource}
\end{figure}

At 1611 days is a large drop in the measured flux density. This can also be seen in Figure \ref{fig:checksource}, where the measured flux density of a background check source in the field is compared with the measured flux density of AT2019azh. We see that at 1611 days the flux density of both the background check source and AT2019azh decrease at 5.5 and 9\,GHz, suggesting an offset or error between this and other epochs. To reduce the effect this epoch has on our fitting we apply a correction to each frequency measured in the epoch by taking the weighted average of the flux density of the background source at other epochs and calculating a scaling factor from comparison with the background source in the issue epoch and applying it to the flux density of AT2019azh. While this correction is rather inaccurate at a number of frequencies due to the variability of the background source, the increased uncertainties are consistent with the light curve and reduce the impact of the affected epoch on the fitting.

\newpage
\section{Flux densities}

\begin{longtable}{ccccccc}
\caption{Measured flux densities at each epoch for both VLA and GMRT.}\\
    Date & $\delta$t days & Instrument & Band & Frequency (GHz) & Flux density($\mu$Jy) & Uncertainty ($\mu$Jy)\\
    \endhead
    \endfoot
      \hline
      20-04-2022  &  1167  &  VLA &  S  &  2.24  & 1550 & 48 \\
                  &       &       &    S  &  2.76 & 1304 & 30 \\
                  &       &       &    S  &  3.24 & 1155 & 25 \\
                  &       &       &    S  &  3.76 & 984 & 24 \\
                  &       &       &    X  &  11 & 318 & 18 \\
       05-05-2022 &  1182 &  VLA  &  L  &  1.14 & 2716 & 68 \\
                  &       &       &   L &  1.42 & 2482 & 36 \\
                  &       &       &    L  &  1.68 & 2129 & 47 \\
                  &       &       &    L  &  1.9 & 1947 & 39 \\
                  &       &       &    C  &  4.49 & 853 & 31 \\
                  &       &       &   C   &  5.51 & 736 &  32  \\
                  &       &       &   C   &  6.49 &  590 &  31 \\
                  &       &       &   C &   7.51 &  445  & 32  \\  
       24-07-2022 & 1262 & VLA & S &  2.5  &  1300  &  490   \\
                  &       &       &  S  &  3.5  &  661  &  177  \\  
                  &       &       &  C &   4.49  & 421  &  101  \\
                  &       &       &   C  &  5.51  &  729  & 102  \\
                  &       &       &  C   &  6.49  &  614  &  60  \\
                  &       &       &   C   &  7.51  & 472  &  58   \\
                  &       &       & X    &   9.23   & 392   & 20   \\  
                  &       &       &    X  &  11.13    & 392  & 20   \\  
       21-01-2023 &  1443  &  VLA  &  L  &  1.14  &  3010 &  110   \\  
                  &       &       &  L  &  1.39  &  2209 & 31   \\
                  &       &       &  L  &   1.65 &  1787  & 73  \\  
                  &       &       &  L  &  1.9  &  1620  &  47   \\
                  &       &       &  S  &  2.24  &  1446  &  45 \\
                  &       &       &  S  &  2.76    & 1142  &  21 \\
                  &       &       &  S  &  3.24  &  996  &  19 \\
                  &       &       &  S  &   3.76  &  864  &  19  \\
                  &       &       &  C  &   4.49  &  696  &  51 \\  
                  &       &       &  C  &   5.51  & 649  &  45   \\  
                  &       &       &  C  &  6.49   & 444 &  42  \\  
                  &       &       &  C &   7.51 & 442   & 34    \\  
                  &       &       &  X   & 9   & 369 & 14  \\
                  &       &       &  X  &  11  & 297 &  18   \\  
       08-07-2023 & 1611 & VLA  & L & 1.26 & 1254  & 49 \\
                  &       &       & L & 1.65 &  1223  &  94   \\  
                  &       &       & L  &  1.9  &  1174  & 38  \\
                  &       &       &  S  & 2.24   &  1193  &  47 \\
                  &       &       &  S &  2.76   & 1021  & 31 \\
                  &       &       &  S &   3.24    & 854   & 27    \\  
                  &       &       &   S  &  3.76   &  734  &  34 \\  
                  &       &       &   C  &  4.49   &  602  &  23   \\
                  &       &       &   C    & 5.51  & 422  &  27   \\  
                  &       &       &   C   &  6.49   & 373  &  15  \\  
                  &       &       &   C &  7.51  &  245  &   22  \\  
                  &       &       &   X &   9  &  177  &  22 \\  
                  &       &       &   X &  11  &  124  &  16   \\  
     27 -12-2023  &  1783 &   VLA & S  &  2.25   &  1074 &  67 \\
                  &       &       &   S &  2.75   & 960  & 37  \\
                  &       &       &   S  &  3.24  & 815  &  26 \\
                  &       &       &  S  &  3.75 &  753  & 33  \\
                  &       &       &  C  &  4.49  &  639  & 28    \\  
                  &       &       &  C  &  5.51  &  458  & 28  \\  
                  &       &       & X  & 9  &  336  & 16 \\  
                  &       &       &  X  & 11  &  270  &  19 \\  
       05-07-2024 & 1974 & VLA & L & 1.14 & 1546 & 65 \\
                  &       &       & L & 1.39 & 1498 & 51 \\  
                  &       &       &  L & 1.65 & 1331 & 54 \\  
                  &       &       & L & 1.9 & 1179& 36 \\  
                  &       &       & S & 2.24 & 973 & 37 \\  
                  &       &       & S & 2.76 & 894& 31\\  
                  &       &       &  S  & 3.25  & 753  &25 \\  
                  &       &       &  S & 3.75 & 664 & 26 \\  
                  &       &       &  C  & 4.5  & 579  & 25  \\  
                  &       &       & C &  5.5 & 519 &30  \\  
                  &       &       & C  & 7 & 385 & 23 \\
                  &       &       & X & 9  & 305 & 18 \\
                  &       &       &  X& 11 & 227 & 17 \\  
       06-01-2025 & 2159 & VLA & L & 1.14 & 1707  & 94 \\
                  &       &       & L  & 1.39  & 1341  & 71\\
                  &       &       & L  & 1.65  &  1192 &70 \\
                  &       &       &  L &  1.9  &  1126 &64 \\
                  &       &       &  S & 2.25  & 1000 &61 \\
                  &       &       & S & 2.75 & 856 &49 \\
                  &       &       & S & 3.25 & 721 & 42 \\
                  &       &       & S & 3.75 & 625 & 38 \\
                  &       &       & C & 4.5 & 540 &34 \\
                  &       &       & C & 5.5 & 471 & 34\\
                  &       &       & C & 6.5 & 370 & 27\\
                  &       &       & C & 7.5 & 299 &24 \\
                  &       &       & X & 9 & 312 & 21\\
                  &       &       & X & 11 & 235 & 18\\
        25-06-2022& 1233  &  uGMRT  &  4   &  0.65  & 3964   &  203  \\     
                  &       &         & 5   &  1.26 & 2487 & 139 \\
        03-01-2023&  1425 & uGMRT & 4 & 0.65 & 3049 & 147 \\
        02-01-2024& 1789  & uGMRT & 4 & 0.65 &3265 &168\\
                  &       &     & 5 & 1.26 & 1559 & 126 \\
       \hline\\
       \label{tab:results}
\end{longtable}

\bibliography{papers}{}
\bibliographystyle{aasjournalv7}

\end{document}